\documentclass[conference]{IEEEtran}
\IEEEoverridecommandlockouts

\usepackage{cite}
\usepackage{amsmath,amssymb,amsfonts}
\usepackage{algorithmic}
\usepackage{graphicx}
\usepackage{textcomp}
\usepackage{xcolor}
\usepackage{url}
\usepackage{hyperref}
\usepackage{booktabs}
\usepackage{array}
\newcolumntype{B}[1]{>{\bfseries}p{#1}}
\newcolumntype{L}[1]{>{\raggedright\arraybackslash}p{#1}}

\newif\ifwithComments
\withCommentstrue

\usepackage{xcolor}
\usepackage{xspace}

\ifwithComments
  \newcommand{\dmg}[1]{{\color{red}dmg says: #1}\xspace}
  \newcommand{\dmgself}[1]{{\color{red}dmg note to himself: #1}\xspace}
  \newcommand{\grex}[1]{{\color{purple}grex says: #1}\xspace}
\else
  \newcommand{\dmg}[1]{}
  \newcommand{\dmgself}[1]{}
  \newcommand{\grex}[1]{}
\fi

\newcommand{\rqone}{\emph How are OSS projects adapting their contribution policies to govern AI-mediated contributions?\xspace}

\newcommand{\rqtwo}{\emph What governance dimensions emerge from these policies, and how can they be organized?\xspace}

\usepackage{framed}
\usepackage{xparse}

\newsavebox{\fminipagebox}
\NewDocumentEnvironment{hassanbox}{m O{\fboxsep}}
 {\par\kern#2\noindent\begin{lrbox}{\fminipagebox}
  \begin{minipage}{#1}\ignorespaces}
 {\end{minipage}\end{lrbox}\makebox[#1]{\kern\dimexpr-\fboxsep-\fboxrule\relax
    \fbox{\usebox{\fminipagebox}}\kern\dimexpr-\fboxsep-\fboxrule\relax
  }\par\kern#2
 }

\def\BibTeX{{\rm B\kern-.05em{\sc i\kern-.025em b}\kern-.08em
    T\kern-.1667em\lower.7ex\hbox{E}\kern-.125emX}}

\begin{document}

\title{You can contribute if you... 
An Empirical Framework of AI Contribution Policies in OSS}

\author{
\IEEEauthorblockN{Gregorio Robles}
\IEEEauthorblockA{\textit{Universidad Rey Juan Carlos} \\
Madrid, Spain \\
gregorio.robles@urjc.es}
\and
\IEEEauthorblockN{Daniel German}
\IEEEauthorblockA{\textit{Department of Computer Science} \\
  \textit{University of Victoria} \\
  Victoria, BC, Canada \\
  dmg@uvic.ca}
}

\maketitle

\begin{abstract}

Artificial intelligence is reshaping open source software (OSS) contribution by lowering the cost of producing code,
documentation, issue reports, and review interactions. This creates opportunities for broader participation, but also
disrupts how maintainers assess contributor effort, competence, and accountability. In response, OSS projects
are beginning to regulate AI-mediated contribution through contribution guidelines and other project
documentation. This paper presents an empirical study of these emerging policies.
We analyze project policies on AI-mediated contributions by evaluating their underlying rationales, rules, and expectations.
Our analysis shows that these policies seek to protect scarce maintainer attention, preserve
accountability, sustain meaningful review interactions, address legal and quality concerns, and maintain pathways for newcomer learning. 
Based on these findings, we introduce the AI Contribution Governance Framework, which organizes
recurring concerns and governance mechanisms across projects. The framework helps OSS communities develop AI
contribution policies and provides researchers with a vocabulary for studying how AI is changing collaborative
software production.

\end{abstract}

\begin{IEEEkeywords}
open source, artificial intelligence, software engineering
\end{IEEEkeywords}

\section{Introduction}
\label{sec:introduction}

Open Source Software (OSS) is commonly developed through public repositories in which people outside the maintainer
group can propose and submit changes to a project. This form of participation allows projects to receive bug fixes, new
functionality, documentation improvements, and other forms of work from a broad community of users and
developers. Maintainers remain responsible for deciding which proposed changes become part of the project, which require
revision, and which are rejected. Contribution policies support these decisions by stating how contributions should be
submitted, what kinds of participation are acceptable, and what responsibilities contributors accept when they submit
work
\cite{crowston-2005-social-struc,bosu-2014-impac-oss,gaughan-2025-readme-contributing,falcucci-2025-contribution-guidelines-testing}.

As described in \cite{feng-2026-chart-uncer-water}, generative AI is changing OSS development.
We use the term \emph{AI-mediated contribution} to refer to contributions whose production, explanation, revision, review, or evaluation is materially shaped by generative or agentic AI systems.
Before widespread use of AI-mediated development, creating a contribution usually required an investment in time,
effort, and project-specific engagement that many attempts never became submissions. This acted as a filter that reduced
their volume and made submission itself evidence that the contributor had invested enough effort to make the
contribution worth reviewing, and that, even if the contribution was weak, nurturing the contributor might be a good
long-term investment \cite{balali-2018-newcom-barrier}. The
use of AI lowers the cost of producing contributions, increasing their volume, and weakens the signals that allowed
maintainers to distinguish promising newcomers from low-commitment contributors.

Maintainers must assess whether the proposed contribution is correct, useful, maintainable, and worth carrying forward
as part of the project (including whether and how AI was used).  Review is already shaped by limited maintainer
attention, contributor reputation, and the costs of evaluating work from peripheral or unfamiliar contributors
\cite{bosu-2014-impac-oss,guizani2021long,guizani-2021-long-road-ahead}. Recent empirical work shows that
pull requests created with the help of AI are already observable at substantial scale in GitHub repositories \cite{li-2026-aidev},
while studies of AI coding tools identify risks involving quality assurance, code complexity, and tool failure modes
\cite{he2026speed,zhang-2026-engin-pitfal,wang-2025-agent-softw-engin}.

While the use of AI can have harmful consequences for project quality and maintainer workload, it can also benefit
OSS. It can help contributors understand unfamiliar code, express ideas more clearly, and prepare submissions that would
otherwise be difficult or time consuming for them to produce; and it can help maintainers in the triage and evaluation
of submissions. This is already being observed \cite{li-2026-aidev}. The governance problem for OSS
projects is therefore how projects can preserve accountability,
maintainability, and community formation while allowing useful forms of AI mediation.

As AI-mediated contribution makes external work harder to evaluate and incorporate, projects are beginning to update their contribution policies in different ways.
Recent empirical work \cite{hora2026aipolicy} measured this emerging policy landscape, showing that many AI contribution policies address permission, disclosure, and human-in-the-loop requirements. Furthermore, the mentioned study demonstrates that while the topic is widely debated nowadays within the OSS community, project positioning regarding AI-mediated contributions is still nascent. This is evidenced by the fact that only 118 of the 1,000 most active GitHub repositories that were studied have established some policy on this issue, many of which remain notably vague.
We build on this line of work by analyzing the governance rationales encoded in these policies.

This paper presents an empirical study of OSS project policies related to AI-mediated contributions.
It is guided by the following two research questions:

\begin{itemize}
\item \textbf{RQ1}: \rqone
\item \textbf{RQ2}: \rqtwo
\end{itemize}

RQ1 maps the emerging OSS policy landscape by examining the rules and expectations that projects introduce for
AI-mediated contribution. RQ2 builds on this analysis to derive an AI Contribution Governance Framework that organizes
the main concerns projects address when adapting contribution policy to AI.

This paper makes two primary contributions. First, it provides an empirical analysis of how open-source software (OSS) projects manage AI-mediated contributions, categorizing the various rationales behind these policies. 
Second, it introduces the AI Contribution Governance Framework as a tool for explaining
existing policies and for guiding projects that are creating or revising policies to address AI use in their own
communities.

\section{Related Work}
\label{sec:related}

OSS contribution is shaped by a persistent tension between community growth and maintainer capacity. Projects depend on
new contributors to sustain the community, replace lost knowledge, and develop future maintainers, but each contribution
also requires maintainers to evaluate, integrate, and support work under limited time and attention. Prior work has
shown that OSS projects fail, become abandoned, or struggle to survive for reasons that include maintainer overload,
insufficient contributor retention, and loss of project knowledge \cite{coelho-2017-why, avelino2019abandonment,
  hata-2015-charac-sustain,rashid-2017-explor-knowl}. Even successful projects operate under constraints: need for heterogeneous contributors, uneven support processes, and ongoing coordination challenges~\cite{guizani2021long}, while pull requests may be abandoned when contributors and maintainers fail to converge on a
review outcome~\cite{li-2022-are-you}. Retaining
contributors and sustaining participation require active governance work, not merely technical infrastructure~\cite{feng-2026-addres-oss}.

Studies of GitHub pull requests show that both social and technical factors shape
contribution evaluation~\cite{rigby-2014-peer-review,tsay-2014-influen-social-technical,german-2018-was}, and code review outcomes are influenced by developer
reputation~\cite{bosu-2014-impac-oss}, trust and security expectations \cite{wermke-2022-commit-trust}, and the social
structure of OSS communities~\cite{crowston-2005-social-struc} and the reviewing process enhances
knowledge sharing~\cite{rigby-2013-convergent-review}. Other work has examined how
core developers can be identified through observable project activity~\cite{bock-2023-autom-core} and how social
interactions shape developer initiation in OSS foundations~\cite{gharehyazie-2014-devel-initiat}.

A related body of work examines newcomers, mentoring, and peripheral participation. Newcomers face both technical and
social barriers when making their first contributions~\cite{steinmacher-2015-social-barrier,
  steinmacher-2018-overc-social, balali-2018-newcom-barrier}, and projects have developed mechanisms to help them select
appropriate tasks and overcome entry barriers~\cite{steinmacher-2015-under-suppor, steinmacher-2016-overc}. ``Good first
issue'' labels are one such mechanism: they identify tasks that are suitable for onboarding and learning
~\cite{tan-2020-github}. Mentoring programs such as Google Summer of Code show challenges of
onboarding new contributors~\cite{silva-2020,tan-2023-under-mentor, feng-2024-guidin-way,
  feng-2025-multif-natur}. Studies of one-time, casual, episodic, and peripheral contributors show that OSS projects
depend on multiple contributor types whose motivations and future trajectories differ ~\cite{lee2017understanding,
  lee-2017-are-one, pinto-2016-more-common, barcomb-2020-uncov-perip, barcomb-2022-manag-episod,
  trinkenreich-2020-hidden-figur,gerosa2021shifting}. Recent work further examines how first contributions can lead, or fail to lead, to
longer-term contributors~\cite{turzo-2025-from-first}.

OSS communities govern participation through documented rules and norms. Prior research has studied the
evolution of OSS governance practices~\cite{noni-2013-evolut-oss-gover}, the adoption and effects of codes of conduct
\cite{tourani-2017-code, singh-2021-codes-conduc, frluckaj-2024-codes-conduc}, and the ways codes of conduct change
community engagement over time \cite{sun2026beyond}. Research has shown that projects use these documents to communicate procedures, expectations,
and norms for participation
\cite{gaughan-2025-readme-contributing,falcucci-2025-contribution-guidelines-testing}.
Contributors
may be driven by ideology, learning, reputation, employment, reciprocity, or commitment to a cause
\cite{rossi2006decoding, chou2011understanding, sharma-2022-motiv-hygien, taylor-2021-for-love,
  jahn-2025-blend-code-cause, coelho-2018-why-floss}.

OSS communities govern participation through explicit documents, rules, and norms. Research has studied the
evolution of OSS governance practices \cite{noni-2013-evolut-oss-gover}, the adoption and effects of codes of conduct
\cite{tourani-2017-code, singh-2021-codes-conduc, frluckaj-2024-codes-conduc}, and the ways codes of conduct change
community engagement over time \cite{sun2026beyond}.  At the same time, governance operates within communities whose
members contribute for varied reasons, including ideology, learning, reputation, employment, reciprocity, and commitment
to a cause \cite{rossi2006decoding, chou2011understanding, sharma-2022-motiv-hygien, taylor-2021-for-love,
  jahn-2025-blend-code-cause, coelho-2018-why-floss}.

Recent work has begun to examine generative AI and software engineering more directly. AI coding tools can increase
short-term development velocity while introducing longer-term quality costs \cite{he2026speed}, and AI coding systems
exhibit recurring engineering pitfalls \cite{zhang-2026-engin-pitfal}. At the same time, AI agents are becoming
increasingly capable participants in software engineering workflows \cite{wang-2025-agent-softw-engin}, and
agent-authored or agent-assisted pull requests are now observable on GitHub \cite{li-2026-aidev}. Recent empirical work
on AI coding-agent adoption further suggests that these tools may reshape OSS participation by changing the relative
share of human and newcomer activity while increasing the need and depth for review\cite{zhang-2026-augmentation-dilution}.

OSS foundations also frame AI-mediated contribution as a governance problem for open,
peer-produced software communities.  Linux Foundation Research identifies maintainer capacity as a central constraint in
OSS, emphasizing that project growth must remain aligned with the coordination, review, and community-management work
maintainers can sustain \cite{salkever2023openSourceMaintainers,gerosa2025stateGlobalOpenSource}.  Its generative AI
guidance connects these concerns to disclosure, provenance, licensing, and human responsibility
\cite{linuxfoundationGenerativeAI}. Mozilla and Wikimedia extend this framing to public-interest and volunteer
peer-production contexts, emphasizing accountable AI systems, human judgment, community oversight, transparency, and
monitoring
\cite{surman2024acceleratingTrustworthyAI,mozillaTrustworthyAI,albon2025wikipediaHumansFirst,wikimediaAIGuidelines}.

Fent et al. \cite{feng-2026-chart-uncer-water} describe how generative AI is disrupting OSS emphasizing how 
OSS communities should discuss and address the different aspects of this disruption.
Closest to our work, Hora and Robbes studied how OSS contribution guidelines address generative AI, focusing on project stances
toward AI-generated contributions, disclosure requirements, and human-in-the-loop responsibility
\cite{hora2026aipolicy}. They show how OSS projects specify rules for generative AI use in contribution processes. Our study
builds directly on this line of work by including their dataset in our analysis of AI contribution policies
and by
examining AI contribution policies as emerging governance responses to the capacity, signaling, onboarding, and
community-management problems that already structure OSS participation

\section{Methodology}
\label{sec:methodology}

To investigate how OSS projects are addressing the review of AI-mediated contributions, we have followed a dual approach inspired by the two main schools of thought in
the philosophy of law: legal positivism and natural law (iusnaturalism).  Legal positivism contends that law is an
exclusively human and social creation, establishing a strict separation between law and morality. According to this
current, the validity of a law does not depend on its level of justice or its alignment with universal ethical values,
but solely and exclusively on its enactment by the competent authority following the formal procedures established by
the State. In other words, a law is valid and must be complied with simply because it has been issued by the official
legislator, regardless of whether its content is morally good or bad.

On the other hand, natural law theory (or iusnaturalism) maintains that there are universal, immutable moral principles
and standards of justice inherent to human nature that are prior and superior to any written regulation. From this
perspective, law enacted by individuals or States is only truly valid, legitimate, and binding if it respects and aligns
with these fundamental ethical values. In short, for a natural law theorist, a law that is profoundly unjust loses its
moral essence and is not considered true law.

This research is grounded in two research questions, one for each philosophical stance:

\begin{itemize}
\item \textbf{RQ1}: \rqone The goal of this research question is to identify the main concerns that OSS projects have and
  organize them into themes.
\item \textbf{RQ2}: \rqtwo We aim to create a framework that is derived from the basic tenants surrounding
  participation in OSS that is capable of capturing and organizing the concerns found in RQ1.
  This framework should result in a set of principles that OSS projects should use to ground the discussion and enactment of
  policies regarding AI-mediated contributions by its members.
\end{itemize}

RQ1 was conducted on the contribution guidelines that open-source software projects publish on their websites and repositories, specifically regarding how to collaborate using AI. This represents the positivist dimension, which methodologically relies on an empirical analysis of artifacts within software repositories. 

The following steps were taken to analyze the existing rules in OSS projects. First, lists of projects that have published such rules were identified, and the webpage with the policy regarding AI-mediated contributions downloaded. We found two sources: (i) an existing list of policies by OSS projects about how to engage with AI-generated contributions, and (ii) a replication package from a pre-print short paper \cite{hora2026aipolicy} that identified projects with such policies (their work focused on the categorization of projects into AI is welcomed, permitted and discouraged, and what requirements for disclosure they impose on contributors).

Next, we attempted to identify the stance (permitted, regulated, prohibited) of the AI project and other regulations included in these guidelines, determining various types of characteristics. These additional characteristics included in the analysis (such as disclosure requirements, the scope, human responsibility) serve to clarify the rationale behind each project's stance and may offer a more comprehensive view of the landscape. This process was performed manually for twelve randomly selected projects until saturation was reached, as no additional characteristics were found. This allowed us to construct an analysis template. This analysis template (shown in Table~\ref{tab:ai_policy_criteria}) was then transformed into a prompt that was passed for every project, together with the webpage with the policy, to Claude Code (using Opus 4.7). The prompt that has been used is as follows \emph{``I am trying to analyze AI policies for following OSS projects. Inspect the content of the files with information on how to contribute and create a table based on following criteria:''}, together with the analysis template.

\begin{table}[htbp]
\centering
\caption{AI Policy Analysis Criteria.}
\label{tab:ai_policy_criteria}
\small
\setlength{\tabcolsep}{4pt}
\begin{tabular}{@{}p{0.32\columnwidth} p{0.65\columnwidth}@{}}
\toprule
\textbf{Category} & \textbf{Possible Values / Options} \\
\midrule
Overall stance & permitted \textbar regulated \textbar prohibited \textbar unclear \\
Disclosure req. & yes \textbar no \textbar recommended \textbar unclear \\
Human resp. & explicit \textbar implicit \textbar absent \\
Scope & everything \textbar code \textbar docs \textbar tests \textbar trans. \textbar issues \textbar props. \\
Quality/testing & yes \textbar no \\
Learning/understanding & yes \textbar no \\
Security/legal & yes \textbar no \\
Authorship/IP & yes \textbar no \\
Sustainability & yes \textbar no \\
Enforcement & rejection \textbar label \textbar review \textbar warning \textbar none \textbar ban \\
Target actor & contributor \textbar maintainer \textbar bot \textbar student \textbar org. \\
Human interactions & explicit \textbar implicit \textbar absent \\
\bottomrule
\vspace{0.5mm}
\end{tabular}
\end{table}

\begin{table*}[t]
\centering
\caption{Sample of coded AI contribution policies (six example projects from the unified dataset; full list in the reproduction package). \textbf{Stance}: Perm.\,=\,Permitted, Regu.\,=\,Regulated, Proh.\,=\,Prohibited. \textbf{Dataset}: HR\,=\,merged Dataset~1+2 (dedicated policy files / CONTRIBUTING files), MW\,=\,melissawm curated list. \textbf{Disclosure}: Y\,=\,Yes, N\,=\,No, R\,=\,Recommended. \textbf{Human Resp.} (human responsibility) and \textbf{H$\leftrightarrow$H} (human--human interaction): E\,=\,Explicit, I\,=\,Implicit, A\,=\,Absent. \textbf{Quality}, \textbf{Learning}, \textbf{Security}, \textbf{IP}, \textbf{Community} (sustainability mentioned): Y\,=\,Yes, N\,=\,No. \textbf{Enforcement}: Rej.\,=\,Rejection, Ban\,=\,Permanent ban, Warn\,=\,Warning, ExtR\,=\,Extra review, None\,=\,No mechanism stated. Scope abbreviations: C\,=\,Code, D\,=\,Docs, T\,=\,Tests, I\,=\,Issues, Tr\,=\,Translations, P\,=\,Proposals, All\,=\,Everything.}
\label{tab:ai-policy-sample6}
\resizebox{\textwidth}{!}{\begin{tabular}{@{}l l l l l l l l l l l l l@{}}
\toprule
\textbf{Project} & \textbf{Dataset} & \textbf{Stance} & \textbf{Disc.} & \textbf{Hum.Resp.} & \textbf{Scope} & \textbf{Qual.} & \textbf{Learn.} & \textbf{Sec.} & \textbf{IP} & \textbf{Comm.} & \textbf{Enforce.} & \textbf{H$\leftrightarrow$H} \\
\midrule
9001/copyparty       & HR & Proh.  & N & E & C, D, I & Y & Y & N & N & Y & Rej.       & E \\
\addlinespace
actualbudget/actual  & HR & Perm.  & R & E & C, I, D & Y & Y & N & N & Y & Rej., Ban  & E \\
\addlinespace
Adwaita               & MW & Proh.  & N & E & C, D    & N & N & N & N & Y & Rej.       & E \\
\addlinespace
agno-agi/agno        & HR & Perm.  & N & I & All     & N & N & N & N & N & None       & A \\
\addlinespace
anomalyco/opencode   & HR & Perm.  & N & I & All     & N & N & N & N & N & None       & A \\
\addlinespace
anuken/mindustry     & HR & Regu.  & N & E & C       & Y & Y & N & N & Y & Rej.       & I \\
\bottomrule
\end{tabular}}
\end{table*}

In addition, at the end of the prompt, we added the following question \emph{``Are there other elements that are worth mentioning?''}, in case there are other elements in the webpage worth noting.

The results were grouped in a single list. Table~\ref{tab:ai-policy-sample6} offers an excerpt (i.e., the first 6 analyzed projects; the full list is available in the reproduction package). To ensure the consistency of our findings, we compared the LLM-generated results with our manual analysis across the 12 selected repositories. While the LLM's output closely aligned with our manual assessments, we identified some discrepancies, particularly regarding project categorization. As a result, we treated the LLM's output as a preliminary baseline and conducted a comprehensive manual verification of every case. When ambiguity arose, typically due to a lack of explicit policy information—we triangulated our findings with the perspectives of the original dataset authors, who also classify projects based on AI usage (though strictly on a binary "Yes/No" basis).

Our study focuses primarily on understanding the specific concerns of OSS communities and the methods they use to address them. The primary value of our study lies in the classification of concerns and the accompanying rationale, rather than in how many projects fall into each group. We do not aim to have a representative dataset of the most popular stances within the OSS community; rather, our objective is to curate a rich and diverse dataset that captures the full spectrum of existing viewpoints, including the ones that are currently in the minority. Consequently, as we categorized the projects, we documented the specific reasons provided for their positions; these were later synthesized to better understand the various perspectives surrounding policies on AI-mediated contributions.

To answer RQ2, we performed an empirically grounded conceptual synthesis of the RQ1 results. We started by theorizing about the core
principles of open-source collaboration and how AI affects them---without considering the existing rules in specific
projects. Our starting point was the many papers that describe OSS collaboration before and after AI (see Section~\ref{sec:related}).

The dataset and the results of RQ1 became a solid foundation for our analysis. We treated these policies and the results of RQ1 as evidence of the practical governance concerns that projects are attempting to manage when regulating AI-mediated contributions. Specifically, we examined how projects rationalize their policy choices and how they implement those choices through restrictions, conditions, exemptions, and procedural mechanisms. Our focus was on identifying recurring concerns, rather than evaluating the correctness of individual policies.

We then expanded this interpretation through a discussion among the authors, identifying gaps not seen in the policies,
and ourselves
asking how AI-mediated contribution changes the different facets of the OSS collaboration and under what conditions
those changes become governance concerns. This step allowed us to move from project-specific rationales to more general
concerns about roles, resources, signals, costs, and responsibilities in AI-mediated contribution.

Finally, we used the identified concerns as the basis for developing principles for acceptable AI-mediated
contribution. This was an iterative synthesis process. We first translated recurring concerns into candidate
acceptability conditions, then compared those conditions across the policy rationales and mechanisms identified in
RQ1. Because our approach is naturalistic, we also considered the contribution process holistically, asking whether the
candidate principles captured the roles, dependencies, and obligations that make OSS contribution and review possible,
even if there not observed in policies under study. Through discussion among the authors, we refined the candidate
principles by merging overlapping ideas, separating concerns that addressed different governance problems, and checking
that each principle was either traceable to the empirical material or necessary to explain the governance relations
revealed by it.  We also revised the organization of the principles to make the resulting framework coherent. A major design requirement was that the framework remain formative rather than prescriptive.

\section{Datasets}
\label{sec:datasets}

We have selected two distinct data sources to gain a more comprehensive understanding of our research topic, as they are inherently complementary.

The \emph{HR dataset} is derived from the replication package of recent work by Hora and Robbes~\cite{hora2026aipolicy}. They analyzed the 1,000 most-starred GitHub repositories that met specific criteria: a minimum of 100 commits, non-fork status, and at least one commit in 2026. Through this analysis, they identified 118 AI policies for AI-mediated contributions.

The \emph{MelissaWM dataset} of Contribution Policies comes from the {\small\texttt{melissawm/open-source-ai-contribution-policies}}  GitHub repository~\cite{mendonca_open_source_ai_contribution_policies}, a community-curated list of 106 projects. 
This list is curated specifically around the discussion of how projects manage AI-mediated contributions and adapt their
contribution policies accordingly and comes with a rich ideological and rationale diversity, including a list of projects that reject AI-mediated contributions.

\begin{figure}
    \centering
    \includegraphics[width=0.45\textwidth]{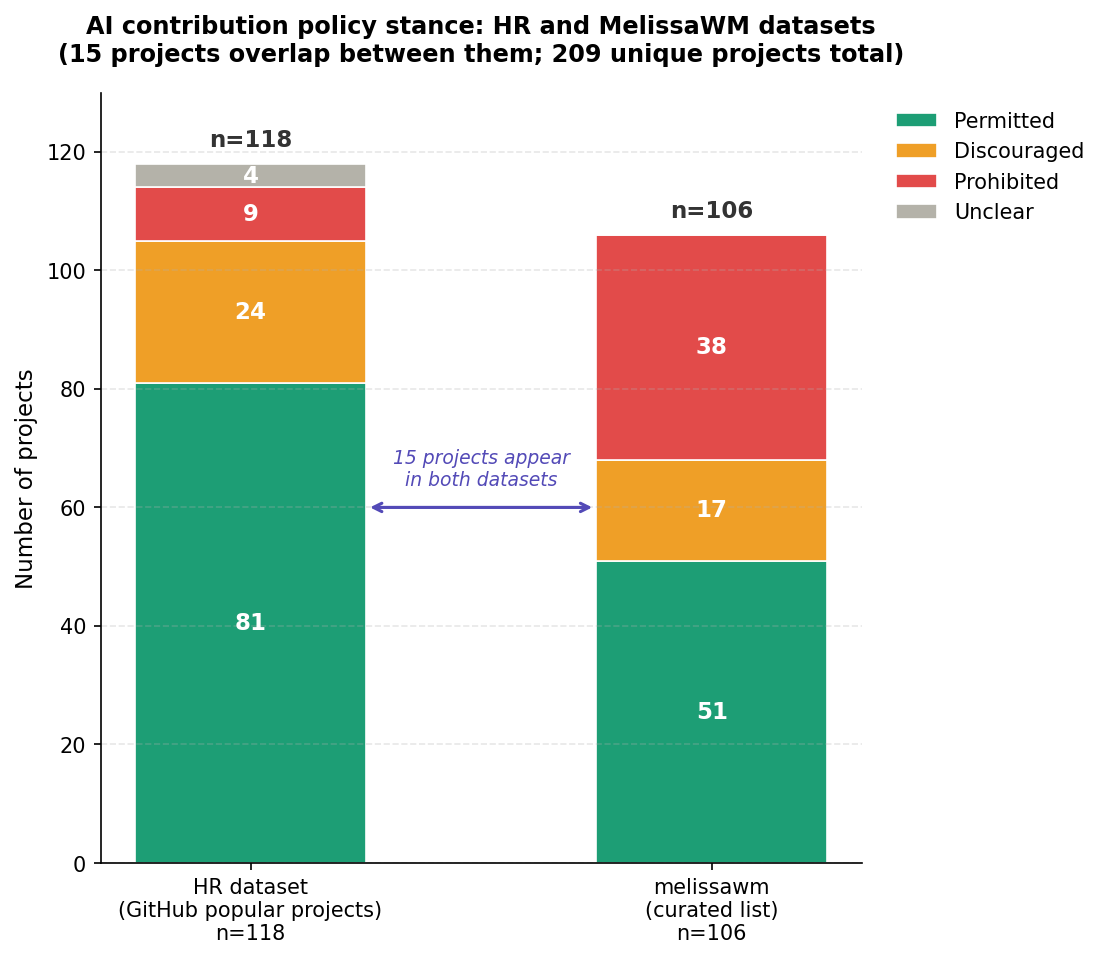}
    \caption{General overview of the number of projects (and their position towards AI-mediated contributions) in the datasets used in this study.}
    \label{fig:overview}
\end{figure}

The datasets have 15 projects in common, resulting in a total of 209 projects.  The analysis of the 209 unique projects
reveals a clear distinction between the two source datasets as can be seen in Figure~\ref{fig:overview}, with the HR
dataset skewing heavily toward AI policies that allow the use of AI (69\%) while the MelissaWM dataset exhibits a more
restrictive stance, featuring a 36\% that prohibit any AI use. We assume that the combined corpus maintains a diverse representation of policy approaches toward AI-mediated content in software development.

\section{Results: RQ1 \rqone}
\label{sec:rationales}

To address the governance challenges introduced by AI-mediated
development in OSS, we establish a classification of current
contribution policies (RQ1). To this end, we have divided this
question into two focused inquiries: a first question (RQ1.1) examines
the specific rationales projects provide for prohibiting, regulating,
or allowing AI-mediated contributions, while a second question (RQ1.2)
investigates whether these contributions are treated
equitably. Specifically, the latter addresses whether policies
differentiate between contributors based on their tenure (new versus
regular), or the nature of the submission (bug fix versus new
feature).

\subsubsection{Reasons by projects prohibiting AI}

\begin{figure}
    \centering
    \includegraphics[width=0.45\textwidth]{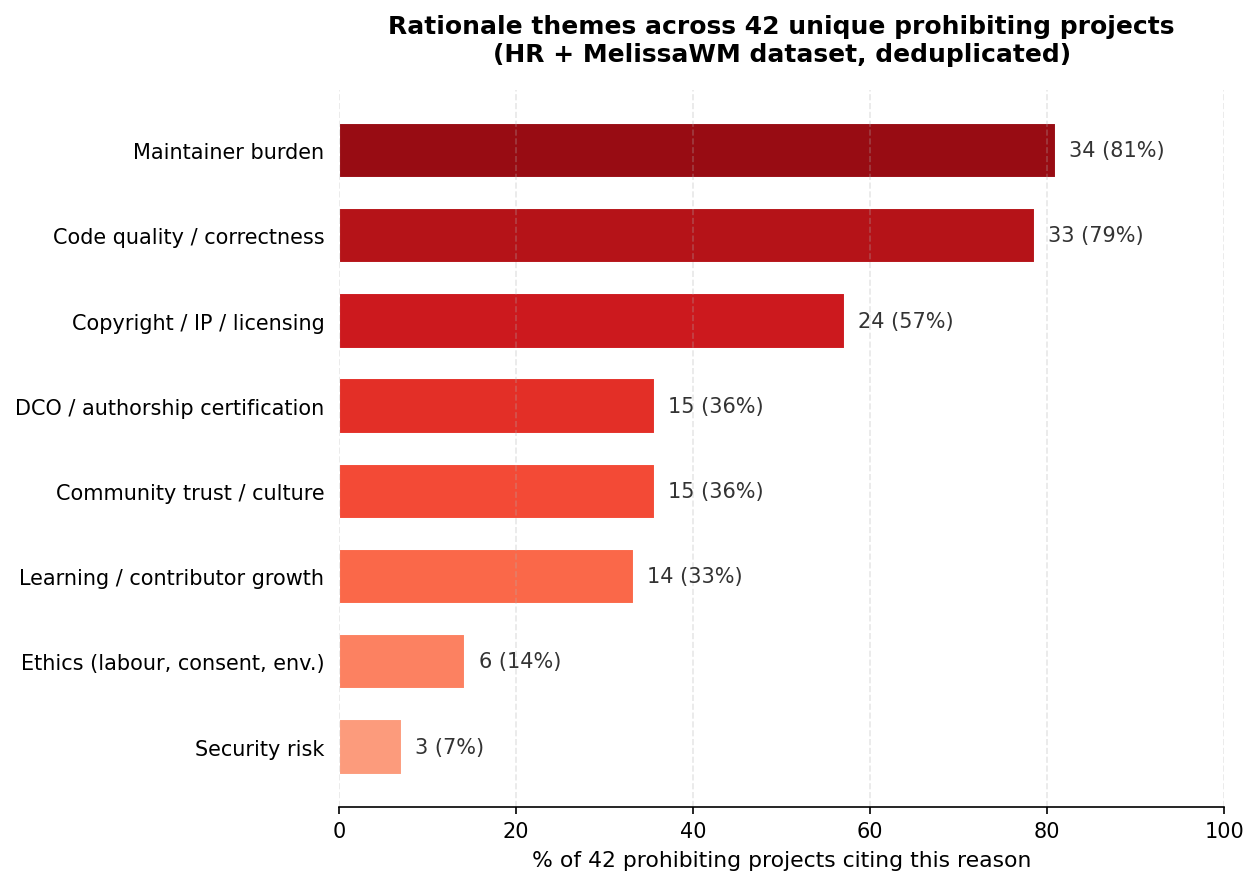}
    \caption{Reasons in projects prohibiting AI-mediated contributions.}
    \label{fig:prohibited-reasons}
\end{figure}

Figure~\ref{fig:prohibited-reasons} offers an overview of the reasons found in projects that prohibit AI.

The dominant and near-universal rationale is the one we have called {\bf Maintainer burden} (34 projects, 81\%). AI eliminates the natural effort-based backpressure that previously limited low-quality submissions---generating large volumes of plausible-looking but broken contributions that consume reviewer time without adding value. Projects citing this reason treat it as an existential threat to sustainability. For example, Ghostty states that \emph{``The rise of agentic programming has eliminated the natural effort-based backpressure that previously limited low-effort contributions.''} and in Zig we can read \emph{``LLM assistance breaks [the relationship between reviewer time and contributor learning] completely… the time the Zig team spends reviewing your work does nothing to help them add new, confident, trustworthy contributors.''}

{\bf Code quality \& correctness} (33 projects, 79\%) comes second, arguing that AI-mediated code is frequently subtly wrong: it passes superficial review, fixes symptoms instead of causes, introduces new bugs, and can only be caught through careful line-by-line reading that defeats the purpose. Projects with high quality standards see AI as incompatible with their quality culture, not merely inconvenient. For example Jellyfin states that \emph{``Pure `vibe coding' will be rejected… If the code looks terrible, it will be rejected as such. You must clean up the mess before submitting.''}

The third most frequent reason refers to {\bf Copyright, IP \& licensing uncertainty} (24 projects \& 57\%), as AI models trained on copyrighted code may reproduce verbatim excerpts of license-incompatible material in their output. This is especially dangerous for projects under copyleft licenses (GPL, MPL) or those operating at legal/regulatory risk. The provenance of generated code is fundamentally unknowable, making it impossible for contributors to honestly certify clean IP. For instance in Asahi Linux it can be read \emph{``There is ample evidence… of this training material including copyrighted material… It is not impossible for [LLMs] to have confidential or leaked material owned by Apple or its vendor partners in their training corpora.''}

The fourth rational is related to {\bf DCO / authorship certification} (15 projects, 36\%), as the Developer Certificate
of Origin (DCO) requires contributors to legally certify they have the right to submit the code under the project's
license. 
Thus, projects that mandate a DCO cannot accept contributions generated by AI that lack proper sign-off.
OpenJDK points this out in their \emph{``Interim Policy on Generative AI''} stating that contributors \emph{``may use generative AI tools privately to help comprehend, debug, and review OpenJDK code.''}, meaning by \emph{privately} that contributors \emph{``may use such tools on [their] own, without contributing the content that they generate.''}

Some projects argue that OSS is a social contract where {\bf Community trust \& culture} (15 projects, 36\%) is fundamental. Contributions are a form of communication between humans---reviewer and author need to be able to have a genuine technical dialogue. When a contributor submits AI-mediated code they do not understand, maintainers are effectively reviewing the AI's output, not engaging with a developer. This is seen as a betrayal of the human relationship at the core of collaborative software. Several projects explicitly frame this as a values issue, not merely a practical one. For instance, SciActive has a \emph{``Human Contribution Policy''} where it states that \emph{``users of the project have an implicit trust that the maintainers of the project understand the code contained within the project's code base''}.

The next most frequent reason given is {\bf Learning \& contributor development} (14 projects, 34\%). These projects state
that OSS contribution is itself a learning process: a pathway through which developers develop expertise, build
relationships with maintainers, and eventually become maintainers themselves. AI short-circuits this
entirely. Contributors who cannot explain their code cannot grow, cannot respond to feedback, and cannot be trusted with
greater responsibility. Several projects frame human learning not just as a side benefit but as the core justification
for accepting contributions at all. For instance, XScreenSaver's maintainer states that \emph{``[n]o contributions built with, or mediated by, LLMs or any kind of ``generative AI'' tools will be considered. If you didn't bother writing it, I'm not going to bother reading it. XScreenSaver is art by humans for humans.''}

The most politically explicit objections to AI are grouped under {\bf Ethics: labour exploitation,
  consent \& environment} (8 projects, 19\%). Projects in this category explicitly oppose using AI tools as a matter of
ethical and environmental responsibility, independent of the quality or legality of the output. This objection is
grounded in several reasons, such as: training data of
LLMs is obtained without creator consent; reliance on exploited low-wage annotators; exploitation of the commons,
including OSS; and the cost of AI inference and training in terms of vast amounts of quantities of energy, water, and
hardware. This category is a political position that makes AI contributions unacceptable regardless of their technical quality. The postmarketOS project exemplifies this concern: \emph{``Their models are built in part with heavily exploited workers in unacceptable working conditions, usually without consent from or compensation for creators of the source material.''} and \emph{``AI tools require an unreasonable amount of energy and water to build and operate… These are harms that we do not want to perpetuate, even if only indirectly.''}

Finally, three projects (7\%) specifically cite {\bf Security} as a primary concern, noting that AI-mediated code frequently contains subtle vulnerabilities capable of bypassing superficial review processes.
This is especially an objection for the security-critical projects in our study 
(e.g., operating systems, password managers); the security risk is unacceptable for them regardless of quality. postmarketOS exemplifies this argument: \emph{``authors making use of generative AI tools oftentimes increase the maintainer burden and cause further problems by not ensuring the following: [...] Making sure their patches are unlikely to introduce bugs, especially security bugs.''}

\subsubsection{Reasons by projects regulating AI}

\begin{figure}
    \centering
    \includegraphics[width=0.45\textwidth]{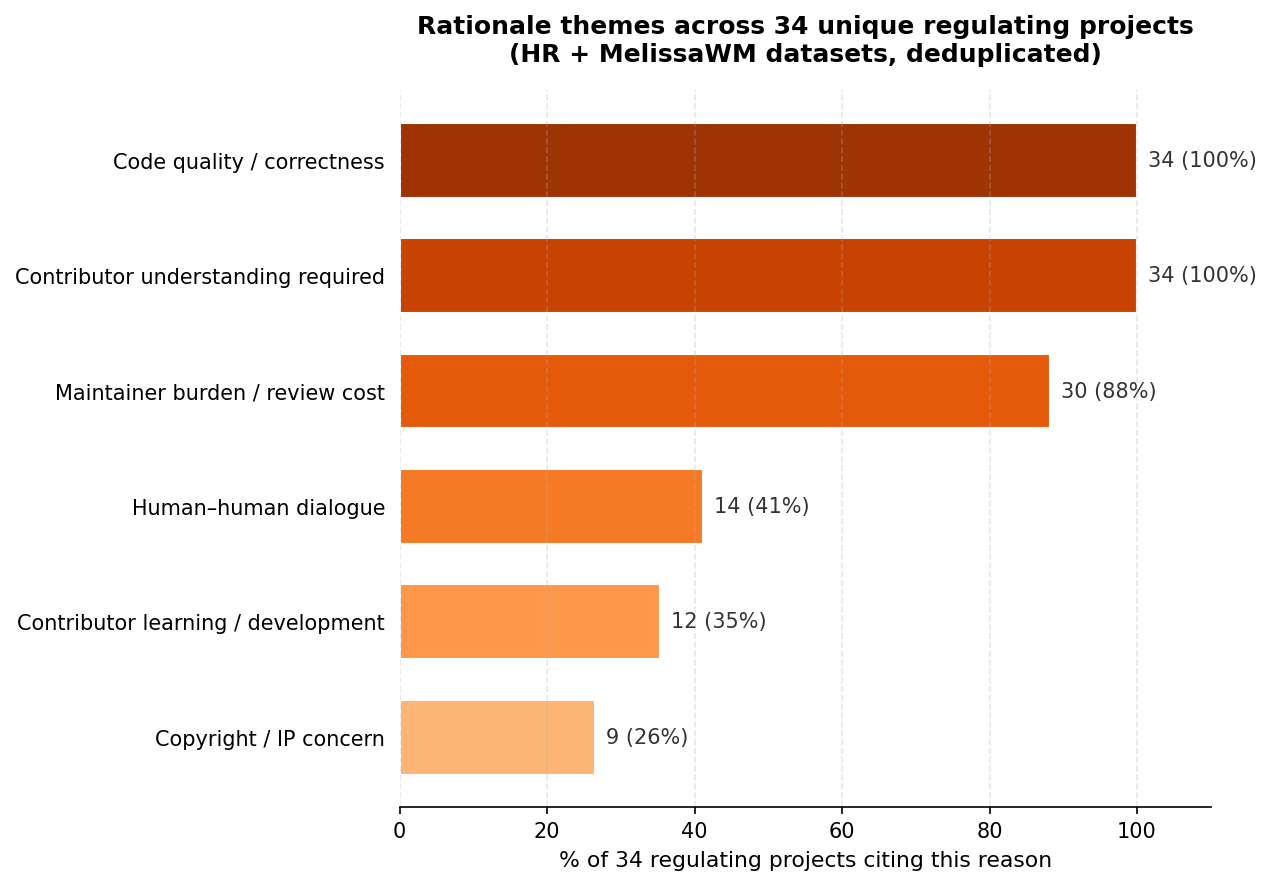}
    \caption{Reasons in projects regulating AI-mediated contributions.}
    \label{fig:discouraged-reasons}
\end{figure}

Figure~\ref{fig:discouraged-reasons} offers an overview of the reasons found in projects that regulate AI.

{\bf Code quality/correctness} and {\bf Contributor understanding required} (34 projects, 100\%) are rationales present in every project that regulate AI contributions, and the one that distinguishes ``regulated'' from ``prohibited'' most clearly. The objection is not to AI tools per se, but to contributors who use them without being able to explain, defend, or take ownership of the output. The framing is consistently about the contributor failing to understand---not about the AI being untrustworthy in the abstract. This makes the bar conditional: AI is tolerated if understanding accompanies it, rejected if it does not. The badlogic/pi-mono project states it as follows: \emph{``Using AI to write code is fine; submitting AI-generated slop without understanding it is not. Human must fully understand and be able to explain changes.''}

The {\bf Maintainer burden \& review cost} reason (30 projects, 88\%) is present in nearly all regulated projects, but it
is expressed differently than in the prohibited group, with a more pragmatic rather than alarmed tone. Where prohibited projects describe a flood that
threatens the project's survival, regulated projects more often describe a more specific review asymmetry: AI lowers the
cost of producing plausible pull requests, while the cost of evaluating, integrating, and maintaining them remains with
maintainers. A major concern is that AI-mediated contributions may
make it harder to tell whether the contributor understands the change and thus whether the submission is worth the required
review effort. FastAPI expresses this concern as follows: \emph{``Do not send
  low-effort or AI-generated PRs. Accounts may be blocked for repeated violations. Human must understand all submitted
  code.''}

A recurring concern in the ``regulated'' group is the degradation of the {\bf Human–human interaction \&
  community dialogue} (14 projects, 41\%) when AI mediates the communication. When a contributor uses AI to draft the
code, the PR description and the responses to review comments, the maintainer is no longer talking to a person but to an
AI that is impersonating her. Several policies state that PR descriptions and review responses must be written by the
contributor themselves (while AI generated code is acceptable). Jellyfin states that \emph{``LLM output is
  expressly prohibited for any direct communication, including issues or comments, feature requests, pull request bodies
  or comments, forum/chat posts.''}.

Several projects that regulate the use of AI express concern that AI bypasses the learning loop, the {\bf Contributor
  learning \& development} group, which is implicit in most projects but explicit only in 12. Contributors who use AI to
generate code they cannot explain are not developing the skills that make future contributions valuable, and they are consuming maintainer time that would otherwise go to genuine mentorship. Learning is framed in the ``regulated'' group as an important outcome of understanding and getting integrated in a project. The ``good first issue'' protection seen in LLVM, similarly found in other projects, reserves beginner-friendly issues as genuine learning opportunities, is a direct expression of this concern: \emph{``AI tools must not be used to fix GitHub issues labelled good first issue. These issues are generally not urgent, and are intended to be learning opportunities for new contributors to get familiar with the codebase.''}

In the regulated group {\bf Copyright \& IP concern}  is also present (9 projects, 26\%), but it is typically expressed as a responsibility placed on the contributor (``you are accountable for the IP status of what you submit'') rather than as a reason to ban AI outright. Projects tend to address copyright implicitly through their disclosure requirements. attrs puts it this way: \emph{``Copyright must be owned line-by-line. No LLM bots in Co-authored-by.''}

\subsubsection{Reasons by projects allowing AI}

Unlike projects that prohibit or regulate AI-mediated contributions, which typically provide a rationale for their stance, we found no such justification in projects that permit AI usage. These projects focus, if at all, on procedural requirements for contributing, primarily emphasizing disclosure, testing, understanding, and human-to-human interaction.  

\subsection{Are Contributions treated Equally?}
\label{sec:contributions-equally}

As policies often fluctuate based on a multifaceted set of criteria, we would like to investigate what factors are
explicitly mentioned in the guidelines, in particular, (i) the contributor's established history and reputation within
the project, and (ii) the nature of the pull request: whether it addresses critical bug fixes or introduces new features.

\subsubsection{New vs. regular contributors}

Only a small number of projects explicitly distinguish between new and regular contributors (17 projects, 8\%), The clearest example can be found in Zig's ``Contributor
Poker'' essay\footnote{\url{https://kristoff.it/blog/contributor-poker-and-ai/}}. Although Zig adopts a universal ban
rather than a role-specific rule, it explains why AI-mediated contributions are especially problematic for newcomers:
first-PR review is an investment in a possible future contributor. AI-mediated
first PRs consume that investment while weakening the signal that would make it worthwhile. Zig is therefore distinctive
because it articulates the importance of the new-contributor dimension even though it does not formalize that
distinction in its policy.

Other projects operationalize the distinction through more specific mechanisms. Five projects, including Ghostty and
Polars, allow maintainer exemptions for trusted contributors, which means that the strict rules for AI usage apply only to outside contributions; maintainers are exempt from these rules and may use AI tools at their discretion as they have proven themselves trustworthy to apply good judgment. Six projects,
including ESLint and go-delve, use an accepted-issue gate to limit contributions to known or pre-discussed work, with
substantial overlap between this category and maintainer exemptions. Six projects, including LLVM and Firefox, protect
good-first issues by reserving them for newcomers. Three projects, including CPython and CCExtractor, apply stricter
rules to mentorship-programme contributors such as GSoC participants. Finally, four projects, including the  Linux Kernel and curl, rely on implicit trust: maintainers acknowledge differentiated expectations in practice, even when these distinctions are not formalized in the official policy.

\subsubsection{Bug fix vs. new feature}

Virtually no policy distinguishes between a bug fix (including addressing an open issue) and a new feature.

DuckDB is the sole project in the entire 209 set that draws the line explicitly by name. Its logic, that bug fixes have a verifiable ground truth while features require design judgement, is the most coherent articulation of why contribution type could matter, although the final decision (\emph{``may be accepted''}) puts it at the maintainer discretion, not as a rule.

The most relevant finding is the inversion in the bug report dimension: projects that are permissive about AI-mediated
code are often strictest about AI-mediated bug report text. NumPy, SciPy, SymPy, Matplotlib, Jellyfin, searxng, and
Zulip all permit disclosed AI code but prohibit AI-written issue descriptions. The reason is that code can have
testability gates that separates correct from incorrect; a bug report has no equivalent mechanism. A non-existent bug
looks identical to a real report until someone spends time trying to reproduce it, consuming valuable maintainer effort\footnote{\url{https://curl.se/dev/contribute.html\#on-ai-use-in-curl}}.

\section{Results: RQ2 \rqtwo}
\label{sec:model}

The results of RQ1 show a wide variety of stances regarding the use of AI, both in terms of the problems projects observe and the mechanisms they use to address them. One common aspect they all share, however, is that they respond to the same broad challenge: AI can increase the volume of contributions, lower their quality, weaken their alignment with project needs, and place additional strain on the review process and on the maintainers responsible for it.

\begin{table*}[t]
\centering
\caption{Concerns for adapting OSS contribution policies to AI-mediated contribution}
\label{tab:ai-model-principles}
\scriptsize
\setlength{\tabcolsep}{3pt}
\begin{tabular}{@{}p{0.11\textwidth} p{0.17\textwidth} p{0.38\textwidth} p{0.31\textwidth}@{}}
\hline
\textbf{Dimension} & \textbf{Governance Concern} & \textbf{What changed} & \textbf{Policy adaptation} \\
\hline

Scarcity shift &
Scarcity shifts from coding to reviewing &
AI makes contributions cheaper to produce, but not cheaper to review, integrate, or maintain. &
Protect maintainer attention. \\

  \hline
Contribution value &
Contribution value varies &
AI lowers production cost most for contributions whose correctness and scope are hard to evaluate, without reducing their review cost.&
Set expectations by contribution type. \\

  \hline
Signal reliability &
Effort signals weaken &
AI reduces the effort needed to produce patches, explanations, tests, and revisions. &
Require evidence of understanding and project fit. \\

&
Review dialogue must remain informative &
AI can generate responses and simulate engagement during review. &
Require substantive human ownership of review dialogue. \\
\hline
Contribution debt&
AI creates multiple debts &
AI-generated contributions can appear correct at review while carrying technical, cognitive, and intention debt that surface during maintenance.&
Address technical, cognitive, and intention debt separately. \\

  \hline
Review effort allo\-ca\-tion under AI &
Burden shifting becomes easier &
Submissions can move verification and integration work to maintainers. &
Permit closure, throttling, or pre-discussion when review burden is excessive. \\

&
Maintainers' use of AI&
AI can triage, and evaluate contributions at scale, making it possible
                        to make acceptance decisions without 
                        justification. &
Allow maintainer-side AI with accountable human decisions. \\

&
Project values shape boundaries &
Projects differ in how they value throughput, quality, learning, authorship, and identity. &
State which values the policy protects. \\

  \hline
Contributor development &
Review is stewardship and investment &
Review integrates code, protects project direction, and develops
                                       future contributors and maintainers. &
Preserve review for work with technical, stewardship, or contributor-development value. \\

&
Newcomer pathways are vulnerable &
AI can bypass the learning process that turns newcomers into long-term contributors. &
Protect good-first issues and mentorship contexts. \\

&
Need for future maintainers &
AI may bypass the learning, trust-building, and sense of giving back through which contributors become maintainers. &
Preserve pathways through which contributors earn trust, develop judgment, and assume long-term responsibility.\\

\hline
Autonomous Agentic contributions &
Accountability\ is un\-assig\-na\-ble & 
Contributions can be initiated and submitted without a human directing the specific submission. &
Require a human operator to accept accountability, or reject fully autonomous submissions outright.                                     \\

\hline
\end{tabular}
\end{table*}

Review is the procedure through which maintainers decide whether a contribution
addresses a problem the project regards as worth solving, fits the project's
technical direction, preserves maintainability, and can be incorporated into the
project's long-term evolution.
We model the review process as a capacity-constrained signaling problem in which review time
is an investment under uncertainty. A contributor chooses whether to submit a
change, and a maintainer must decide not only whether to begin reviewing it, but
also whether to spend scarce review effort to decide
whether to accept or reject the contribution. Before AI-mediated
development, the work required to produce a plausible patch and follow through
with review acted as a costly signal. This signal had two parts. First, effort could indicate
contribution quality:
the contributor had done enough work to make acceptance plausible. Second, for
new contributors, effort could indicate contributor potential: even if the patch
was weak, the contributor had shown enough interest to make mentoring
potentially worthwhile. AI weakens both signals by making polished artifacts
cheaper to produce and by allowing an AI operator to generate, not only the
initial contribution, but also the surrounding materials and interaction such as the pull
request text, the justification, the test scaffolding, the responses to review
comments, and revised patches. As the apparent value of the contribution 
become less informative at time of submission, maintainers
face a harder allocation problem, which is further strained by an increase in contributions
due to AI (including PRs and bug reports).

The framework we propose is naturalistic rather than prescriptive. Its first part identifies the concerns surrounding the
governance problem that the review process faces trying to judge and decide whether the expected value of a contribution
exceeds the expected upfront cost of its review (this threshold will vary across projects).  As AI increases submission
volume and weakens the signals that made review worthwhile, maintainers must devote more attention to triage.
Review
effort is justified when it produces at least one of three returns: i) it may improve the project directly by integrating a
valuable contribution, ii) it may protect the project through stewardship of architecture, maintainability, and
direction, or iii) it may also develop project capacity by helping an interested participant become a long-term
contributor
and potentially, a maintainer.

This framework is derived from the structure of the OSS contribution
system: the stakeholders involved, the resources they contribute, the signals they use, the costs they impose on one
another, and the payoffs they seek and informed by the results of RQ1. Table~\ref{tab:ai-model-principles} describes
these concerns, organized around a set of dimensions. Each dimension identifies a major way in which AI-mediated participation changes the contribution
process; each concern identifies an issue that contribution policies should consider and potentially address.

This framing leads to \textbf{three policy adaptations}. \textbf{First, policies should protect
maintainer attention} through scope limits, pre-discussion requirements, throttles, or closure rules for low-return
submissions. \textbf{Second, policies should preserve mentoring
pathways} for newcomers and new maintainers. And \textbf{third, policies should distinguish contribution types}, since bug fixes, new features, refactorings, and
speculative changes differ in expected value and review burden. Their balance is difficult to achieve: a
policy that filters aggressively may protect maintainers while discouraging future contributors, whereas a policy that
remains too open may preserve access while exhausting review capacity.

\begin{table*}[t]
\centering
\caption{Principles for acceptable AI-mediated contribution}
\label{tab:ai-acceptability-principles}
\scriptsize
\setlength{\tabcolsep}{3pt}
\begin{tabular}{@{}p{0.065\textwidth} B{0.15\textwidth} p{0.45\textwidth} p{0.29\textwidth}@{}}
\hline
\textbf{Layer} & \textbf{Principle} & \textbf{Acceptability condition} & \textbf{Governance function} \\
\hline

Project &
Project autonomy &
Each project may define valid and invalid AI-mediated contribution practices. &
Preserves local governance authority. \\
governance &
Interpretive fairness &
AI use should be judged by its effects, not by reflexive pro-AI or anti-AI assumptions. &
Prevents both blanket suspicion and uncritical acceptance. \\

 &
Project risk boundaries &
Any contribution must fall within the project's accepted boundaries
                          for security, uncertainty, maintenance burden, and provenance. &
Guides screening, review and merge decisions.\\
& Inclusive participation & 
Policies must not exclude contributors for whom AI mediation enables participation 
that would otherwise be inaccessible due to technical, linguistic, or knowledge barriers. &
Ensures AI governance preserves rather than narrows the contributor base.\\
&
Autonomous agentic boundary &
Each project explicitly defines whether autonomous agent submissions are permitted, and if so, under what conditions. &
Defines the boundary for autonomous submissions and accountable human oversight.\\
\hline

Contributor  &
Accountability &
The human contributor understands the contribution well enough to explain, revise it defend its rationale and justify
                 relevant AI use; and to respond to review with genuine human judgment: revise or withdraw the submission, and address problems that arise from it.&
Ensures that maintainers engage with a human participant who is responsible and accountable for the contribution.\\

& No burden shifting &
AI use must not shift unreasonable verification, debugging, explanation, provenance, or long term maintenance costs to maintainers. &
Protects scarce review capacity. \\

&  Evidence for acceptance&
The contributor provides enough rationale, evidence, tests, and responsiveness for maintainers to form justified trust in the contribution, even when they do not fully reconstruct or understand every part of it.&
Enables maintainers to make an acceptance, rejection, or revision decision without having to reconstruct the contribution.\\

& Signal integrity &
AI use must not mislead maintainers about effort, understanding, or risk. &
Preserves the informativeness of contribution signals. \\

& Provenance discipline &
The contribution has acceptable authorship, origin, and licensing status. &
Reduces security and licensing risks.\\
  \hline

Maintainer &
Accountability &
Maintainers remain responsible and accountable for review, triage, rejection, and incorporation decisions. &
Keeps the use of maintainer-side AI subordinate to human stewardship. \\

&  Review integrity &
The maintainer directly evaluates the contribution; AI may summarize or flag but acceptance, rejection, and incorporation decisions must reflect the maintainer's own assessment.&
Ensures review decisions reflect genuine human judgment. \\

\hline
\end{tabular}
\end{table*}

Disclosing that AI was used, and how, is useful for transparency, but insufficient as a signal of contributor
ownership. AI changes the value of contribution signals as they can be generated cheaply (using AI). Projects therefore
may need to require signals with higher evidentiary value. These signals help
maintainers determine whether AI-mediated work (if used) is owned by the contributor or instead leaves evaluation, explanation,
and integration work to the project's maintainers.

Maintainer-side AI should also be considered as part of adaptation. Projects may use AI to increase review capacity
through triage, summarization, testing, or analysis. AI can support review, but decisions about what the project
accepts, rejects, or maintains must remain under human stewardship.

The adaptations required for AI-mediated contributions require projects to specify, in their policies, what forms of AI
use are acceptable and under which conditions. We call these conditions
\textbf{Acceptability Principles: role-specific requirements that AI-mediated participation must satisfy to remain
compatible with the project’s contribution regime}. They define the conditions under which AI use remains compatible with
project stewardship, the central purpose of the project’s contribution policies.
These are presented in Table~\ref{tab:ai-acceptability-principles}.

The acceptability principles are organized around the three stakeholders of OSS contribution governance: the project,
the contributor, and the maintainer. The project defines the contribution regime and its boundaries; the contributor
remains responsible for the submitted work; and the maintainer preserves stewardship over review and incorporation
decisions.

Every OSS project serves a particular need and user base, which gives it its own priorities and goals
which shape the project’s stewardship commitments and, in turn, its contribution
policy. \textbf{Acceptable AI use therefore depends on how a project defines valuable contribution, acceptable risk, contributor
responsibility, and maintainer stewardship.} The acceptability principles provide a way to make those project-specific
conditions explicit. Variation in AI contribution policies (as we observed in RQ1) is expected because projects face different stewardship
conditions: every OSS project serves a particular need and user base,
with its own priorities and goals.

\section{Discussion}
\label{sec:discussion}

A higher level view of the results from RQ1 reveal that 42 projects prohibit the use of AI. These projects do so mainly
for three distinct reasons: 18 prioritize code quality to avoid maintenance burdens (i.e., quality first), 16 cite legal
uncertainty surrounding copyright of AI generated code (i.e., avoid the legal void), and 8 reject AI usage as a matter
of ethical principles (i.e., AI negative impact on society and nature). Improvements in AI-generated output, clearer
evidence of its reliability, or greater legal clarity may lead some projects to revisit their bans. However, projects
that reject AI on ethical or community-value grounds may be less likely to change their position solely because the
technology improves.

Another relevant reflection stemming from the results is that ``regulated'' is not a soft version of ``prohibited''; it
has a distinct and internally consistent logic that sets it apart. Prohibited projects state AI-mediated contributions
are categorically unacceptable; regulated projects find AI  acceptable under the condition that the contributor
demonstrates full understanding and responsibility for the contribution (most policies focus only on the contributor).
Consequently, regulated policies are defined by conditionality (e.g., `if you use AI...').

\subsection{Implications for researchers}

Fent et al provide a roadmap of
\cite{feng-2026-chart-uncer-water}. The cite the importance of
studying how Governance in OSS communities is adapting to the use of AI. The first part of our study shows the wide diversity in which OSS projects are addressing this (including complete banning). Because we have studied only projects that have adapted their policies to AI, we do not know if those who have not are also struggling with AI and accept them. Plenty of research is needed in this direction. Our framework can be used as a foundation for further research that studies how OSS communities are adapting their contribution policies to AI.

Contribution policies are living documents. They are shaped by the communities that create them and by the software those communities steward. The projects that are changing their policies in response to AI-mediated contribution are at the forefront of this transition, as they work out how best to adapt their governance practices. However, most OSS projects have not yet adapted their contribution policies to AI, and we do not know whether they are considering doing so, whether they are already managing AI-related problems informally, or whether AI has not yet become a salient governance issue for them. Further research is needed to answer these questions.

\subsection{Implication for practitioners}

AI contribution policies are project-governance instruments through
which OSS communities decide what they value, what burdens they are
willing to accept, what forms of human accountability they require,
and which forms of participation they want to preserve. The AI
Contribution Governance Framework supports this process as a policy
design aid rather than a prescription for a single AI
policy. Communities can use the framework to identify the challenges
AI creates in their specific context and to translate those challenges
into concrete policy mechanisms, procedures, and enforcement
practices. In doing so, the framework helps communities deliberate
about the values their policies should protect and make those values
explicit and operational in their contribution policies.

\section{Threats to Validity}
\label{sec:threats}

Following the main types of threats to validity in Experimental Software Engineering proposed by Wohlin et al.~\cite{Wohlin2012}, the following three types that affect our study are discussed: internal, external and construct threats.

Our analysis is subject to several threats to internal validity stemming from the construction and verification of the dataset itself. While many project policies were verified against primary sources like CONTRIBUTING.md or AI\_POLICY.md files, others lacked formal documentation. In such cases, the most consequential statements often appeared in informal channels, such as blog posts or public statements (e.g., curl and LadybirdBrowser). In addition, many policy texts were remarkably brief or sparse, making it difficult to interpret their intent. The authors conducted a manual verification supported by a triangulation process to compare their findings against the original dataset authors' classifications, identifying those discrepancies that necessitated the reclassification of projects and the refinement of their underlying rationales.

A threat to external validity is that our results should not be interpreted as estimating the prevalence of AI
contribution policies, policy types, or governance mechanisms across the entire OSS universe. Although we combine two
complementary sources—one based on policies identified among the 1,000 most-starred GitHub projects and another based on
a hand-curated collection of publicly visible OSS AI contribution policies—these sources do not support statistical
generalization to all OSS projects. Our findings therefore should not be read as claims about how common particular
policy positions are in OSS as a whole. Instead, the results identify the range of governance issues, policy mechanisms,
and tensions that appear in current AI contribution policies. We use these results as the empirical basis for developing
our framework for acceptable AI-mediated contribution.

The main construction-validity threat is the conceptual leap from policy evidence to framework construction. The goal of
RQ1 is to classify what current OSS AI contribution policies say; thus these policies become the starting point for
answering RQ2: identifying the broader governance problems that AI-mediated contribution creates for OSS projects. This
means that some framework dimensions go beyond the explicit wording of the policies. The risk is that the framework may
either remain too close to the data, becoming a catalog of current policy features, or move too far from the data,
becoming a theoretical account that is not sufficiently grounded in observed policy responses. We mitigate this by
keeping a clear distinction between the policy mechanisms we observed and the governance issues we infer from them, and
by presenting the framework as an empirically grounded conceptual tool that can be revised as OSS AI governance evolves.

\section{Conclusions}
\label{sec:conclusions}

Communities manifest their values in the software they create, while the software itself shapes the values of the community that forms around it. This creates a feedback loop: an OSS project attracts contributors through the software it offers, but once a community forms, its values influence how the software evolves.

In this study, we show that many OSS communities are regulating AI-mediated contribution in widely different ways, including restriction and outright bans. Their rationales also differ. Some policies are pragmatic, focusing on quality, review burden, maintainability, and legal uncertainty. Others are more philosophical, emphasizing authorship, human participation, community identity, or the broader social and environmental effects of AI. This diversity reflects the heterogeneity of OSS: projects serve different purposes, face different constraints, and form different communities.

AI contribution policies should therefore be understood as adaptations to a changed contribution environment. If AI lowers the cost of submitting work, policies must protect the scarce capacity needed to evaluate and absorb it. If AI weakens traditional signals of effort, understanding, and commitment, policies must require stronger evidence of project fit and human accountability. The purpose is not to maximize or minimize AI use, but to preserve the conditions under which review remains a worthwhile investment in both the codebase, and the contributor and maintainer community.

The AI Contribution Governance Framework supports OSS communities in this process. Rather than prescribing a single policy position, the framework helps communities identify the AI-related pressures that matter in their context, connect those pressures to the values they want to protect, and translate those values into concrete policy mechanisms.

Ultimately, as AI reshapes contribution, the central challenge is for the OSS ecosystem to keep adapting without losing the conditions that allow its projects and communities to thrive.

\bibliographystyle{IEEEtran}

\end{document}